# Pyramidal structural defects in erbium silicide thin films

Eu Jin Tan,[1,2] Mathieu Bouville,[1] Dong Zhi Chi,[1,*] Kin Leong Pey,[2] Pooi See Lee,[3] David J. Srolovitz,[4] Chih Hang Tung,[5] and Lei Jun Tang[5]

[1] *Institute of Materials Research Engineering, 3 Research Link, Singapore 117602*

[2] *School of Electrical and Electronic Engineering, Nanyang Technological University, Nanyang Avenue, Singapore 639798*

[3] *School of Materials Engineering, Nanyang Technological University, Nanyang Avenue, Singapore 639798*

[4] *Department of Mechanical and Aerospace Engineering, Princeton University, Princeton NJ, U.S.A.*

[5] *Institute of Microelectronics, 11 Science Park Road, Science Park 2, Singapore 117685*

A new pyramidal structural defect, 5 to 8 μm wide, has been discovered in thin films of epitaxial $ErSi_{2-x}$ formed by annealing thin Er films on Si(001) substrates at temperatures of 500°C to 800°C. Since these defects form even upon annealing in vacuum of TiN-capped films their formation is not due to oxidation. The pyramidal defects are absent when the $ErSi_{2-x}$ forms on amorphous substrates, which suggests that epitaxial strains play an important role in their formation. We propose that these defects form as a result of the separation of the silicide film from the substrate and its buckling in order to relieve the compressive, biaxial epitaxial stresses. Silicon can then diffuse through the silicide or along the interface to fully or partially fill the void between the buckled $ErSi_{2-x}$ film and the substrate.

Keywords: erbium disilicide, rare-earth, epitaxy, rapid thermal annealing

---

[*] corresponding author: dz-chi@imre.a-star.edu.sg

E.J. Tan *et al.*

Rare earth (RE) silicides are of considerable technological interest for potential use as Schottky sources/drains in metal-oxide-semiconductor field-effect transistors (MOSFET) due to their low Schottky barriers (~0.3 eV) to *n*-type Si [1]. Schottky source/drain technology in semiconductor-on-insulator (SOI) has advantages in scaling and fabrication compared with traditional MOSFETs [2]. However, 'pits' or 'pinholes' are common in RE silicide films formed by solid phase reactions of rare earth metals and Si [3–7]. Such defects can degrade the performance of Schottky barrier devices [8]. The pits are of considerable lateral extent and penetrate deeply into the Si substrate relative to the silicide thickness. The shape of these defects is inherited from the substrate structure: triangular on Si(111) and square on Si(001). In this article, we report the observation of a new type of structural defect in $ErSi_{2-x}$ films formed by rapid thermal annealing (RTA) of Er films on Si(001) substrates. This defect has a pyramidal shape with the apex directed away from the substrate. We characterize this pyramidal structural defect and propose a mechanism for its formation. Unlike pits that form because of local depletion of Si atoms [3, 4, 7, 9] the formation of these defects is associated with $ErSi_{2-x}$/Si epitaxial strains.

The $ErSi_{2-x}$ films were grown by sputter depositing Er on *p*-doped Si wafers (5–10 $\Omega$ cm) followed by rapid thermal annealing. The wafers were cleaned using the standard RCA method, then dipped in dilute HF for 1 minute prior to the deposition of a 50 nm thick erbium film at room temperature in a chamber with a base pressure of $3\times10^{-7}$ torr. In some samples, a 15 nm thick TiN layer was deposited *in-situ* on top of the 50 nm Er (to prevent oxidation). This was followed by a rapid thermal anneal for 60 seconds at between 400 and 800°C in a $N_2$ ambient or in vacuum ($1\times10^{-6}$ torr). The 400°C anneals did not produce $ErSi_{2-x}$. Additional wafers were amorphized using Si ion implantation (30 keV, $1\times10^{15}$ ions/cm$^2$) prior to Er deposition in order to determine the role of epitaxy in the defect formation process.

Phases were identified using x-ray diffraction (Bruker D8 GADDS x-ray diffractometer with 2D detector), electron diffraction and secondary ion mass spectroscopy (SIMS). Figure 1(a) shows a typical XRD pattern for an Er film deposited on Si(001) and annealed for 60 s at 500°C (temperatures up to 800°C showed similar results). This figure shows the $\{1\bar{1}00\}$, $\{2\bar{2}00\}$, $\{1\bar{1}01\}$, and $\{1\bar{1}02\}$ peaks of AlB$_2$ hexagonal erbium disilicide. The observation of dominant $\{1\bar{1}00\}$ and $\{2\bar{2}00\}$ peaks in the 2θ scans suggests that ErSi$_{2-x}$ $\{1\bar{1}00\}$ (prism plane) is oriented parallel to the (001) Si substrate surface, i.e., the c-axis of ErSi$_{2-x}$ lies parallel to the substrate. Furthermore, the observation of the two $\{1\bar{1}01\}$ peaks symmetrically displaced from the central axis (containing the $\{1\bar{1}00\}$ and $\{2\bar{2}00\}$ peaks) implies that this is not simply a fiber texture, but that there is a strong in-plane texture as well. The inset in Fig. 1(a) shows a pole figure of the same film. The 4-fold symmetry (rather than the 2-fold symmetry expected if the ErSi$_{2-x}$ were a single crystal) confirms that there are two distinct ErSi$_{2-x}$ epitaxial orientations relative to the Si substrate. That is, the ErSi$_{2-x}$ film contains both regions where the [0001] (i.e., c-axis) is parallel to Si[110] and regions where it is parallel to Si[1$\bar{1}$0], in agreement with previous reports [5, 10, 11]. All films grown on the crystalline Si substrate and annealed between 500 and 800°C with and without TiN capping exhibited epitaxy.

Figure 2(a) is an optical micrograph of a sample annealed at 500°C for one minute. The image shows that the film contains many pyramidal defects. These defects all have nearly square bases and all of the defects share the same orientation. This micrograph, together with the atomic force microscopy (AFM) images in Figs. 2(b) and 2(c), suggests that these defects are square-based pyramids (apex points away from the Si/ErSi$_{2-x}$ interface). Similar observations were made on samples annealed at 500 ≤ T ≤ 800°C. The pyramidal defects are 4–8 μm wide and have a density on the order of $10^5$ cm$^{-2}$. The presence of these defects in all samples, independently of whether the film was capped with TiN or not prior to annealing in vacuum or in $N_2$, indicates that the formation of these defects is not associated with oxidation. To our knowledge, the presence of such pyramidal defects has not previously been reported. On the other hand, earlier reports suggest that pits penetrating into the Si substrate form when ErSi$_{2-x}$ films are grown on Si(001) at low temperature [3, 4, 9].

Figure 3 shows cross-sectional transmission electron micrographs (XTEM) of pyramidal defects like those shown in Fig. 2. In Fig. 3(a), the space between the ErSi$_{2-x}$ film and the original Si substrate surface is filled with Si, as confirmed using electron dispersive spectroscopy. In Fig. 3(b), the intervening space is only partially filled with Si; i.e., there is a void below the pyramid apex that extends from the film to the substrate. In both cases, the silicon between the film and original substrate surface is highly defective. Electron diffraction and high resolution TEM of the defective area (see the inset in Fig. 3(b)) show that the defective region of the Si consists of a dense array of parallel stacking faults on {111} ((a/6)<110> translations).

The pyramidal defects are absent from ErSi$_{2-x}$ films grown on preamorphized (PAI) Si substrates (the amorphous region extended from the surface into the substrate to 45 nm). Figure 1(b) shows that the ErSi$_{2-x}$ film formed on amorphized silicon is a random polycrystal rather than epitaxial as in growth on crystalline Si(001).

The hexagonal ErSi$_{2-x}$ forms on Si(001) with its [0001] axis parallel to Si<110>. The mismatch parallel to the c-axis is –6.5% (compressive) and +1.1% (tensile) perpendicular to the c-axis. The diffraction data and previous reports [11–13] show that the ErSi$_{2-x}$ film consists of alternating grains with the c-axis parallel to [110] and [$\bar{1}$10]. The grain size is typically 50–150 nm [11, 12], which is much smaller than the lateral extent of the pyramidal defects. Therefore, the misfit strain, as measured on a scale larger than several grains is biaxial and compressive, –2.7%.

The proposed pyramidal defect formation mechanism is illustrated in Fig. 4. The film buckles away from the substrate to relieve the very large epitaxial strain energy (Fig. 4(b)). We note that the pyramidal defect density is too low to significantly relieve the total strain energy stored in the film. Such buckling or blistering must originate at a pre-existing defect or region where the bonding between substrate and film is weak. These may either be processing defects or form as the result of plastic deformation. While it is possible for a compressively stressed film to buckle while remaining attached to the substrate, the observations of voids beneath the pyramidal defect (Fig. 3(b)) and the highly defected Si where the void was filled (insets to Fig. 3(b)), both suggest that the film separates from the substrate upon buckling. Further, if the film remained attached to the substrate upon buckling, the density of the pyramidal defects would be much larger than that observed (to relax much of the strain energy stored in the $ErSi_{2-x}$ film).

After the pyramidal defect forms by buckling, Fig. 4(b), the open region between the film and substrate begins to fill with Si, Figs. 4(c)–(e). This filling occurs in order to lessen the surface energies; filling the void below the buckled $ErSi_{2-x}$ film covers part of both the Si free surface and the $ErSi_{2-x}$ free surface. One indication that this filling occurs is the observation that the Si that fills in, under the buckled $ErSi_{2-x}$ film, has a very different defect structure from that in the original Si substrate. That is, this Si contains a high density of {111} stacking faults (see Fig. 3(b)). The Si atoms located at the $ErSi_{2-x}$/Si interface close to the defect diffuse along or near the interface towards the defect. The stoichiometry of $ErSi_{2-x}$ obtained from Rutherford backscattering spectrometry (RBS) indicates that 1 in 6 Si sites is vacant [14] —hence the notation $ErSi_{2-x}$ with x = 1/3. Therefore, we expect that the Si diffusivity through the $ErSi_{2-x}$, perpendicular to the c-axis, is very high (note that $ErSi_{2-x}$ consists of layers of Si and Er perpendicular to the c-axis). When the Si atoms reach the edge of the defect, they can diffuse either on the Si(001) surface or on the free $ErSi_{2-x}$ surface (Fig. 4(c)). As shown in Fig. 4(c), silicon may grow simultaneously from the two surfaces. In the case of Si growing on the Si(001) surface, only one crystal orientation is possible. In contrast, there is considerable evidence that Si formed on free $ErSi_{2-x}$ surfaces may grow with different orientations (see [15]). Eventually, the two re-crystallization fronts meet as shown schematically in Fig. 4(e), resulting in the filled pyramidal defect of Fig. 3(a). Figure 3(b) shows the case where a void remains (i.e., where the two fronts have not yet met) as in Fig. 4(d).

Examination of a large number of the pyramidal defects in the scanning electron microscope shows that many of the defects are cracked (these can also be seen in the optical micrograph in Fig. 2(a)). The cracking, represented in Fig. 4(f), is likely a result of the large tensile stresses formed near the four edges of the pyramid that meet at the apex and the contraction of the $ErSi_{2-x}$ film upon cooling. The coefficient of thermal expansion of erbium silicide is larger than that of silicon hence large differential strains occur on cooling from the annealing temperature to room temperature.

The mechanistic model described above suggests that a balance between the elastic energy stored in the unbuckled film and the surface and interface energies determines the size of the pyramidal defects. Therefore, once the $ErSi_{2-x}$ film forms and buckles on rapid thermal annealing, the size of the pyramidal defects should not increase on further annealing. As a check on the validity of the mechanistic model, we measured the pyramidal defect size in several films, formed by annealing for different times. The results are shown in Fig. 5. Except for very short times (corresponding to incomplete silicide formation), the pyramidal defect size is independent of annealing time —in agreement with the predictions of the mechanistic model.

ErSi$_{2-x}$ films formed by rapid thermal annealing of Er films on Si(001) exhibit a pyramidal structural defect that has not previously been reported. This defect forms in order to relax epitaxial misfit stresses through a buckling process. The bases of the pyramids are approximately square, indicating that the stresses in the film are biaxial. The biaxial nature of the stress is a result of the fact that the hexagonal ErSi$_{2-x}$ forms epitaxially on Si with two distinct variants; with the c-axis of the silicide parallel to Si[110] and Si[1$\bar{1}$0]. The pyramidal defects nucleate at pre-existing defects and do not form at densities large enough to relax most of the epitaxial stresses. The buckling process occurs simultaneously with separation of the film from the substrate below the pyramidal defect. Silicon diffusion, subsequent to buckling can fill in the space between the buckled film and substrate. Depending on the size of the defect and the annealing time, this can lead to complete filling or partial voids inside the pyramidal defects. The defect size does not increase with annealing times that are long compared to that required for the silicide formation reaction to occur.

The authors gratefully acknowledge useful conversations with Dr. Chris Boothroyd and Dr. Foo Yong Lim in the interpretation of TEM and XRD observations. Financial support for this work was provided by Systems on Silicon Manufacturing Co. Pte. Ltd. (SSMC) and by A*STAR, Singapore.

**References:**

[1] M. Jang, J. Oh, S. Maeng, W. Cho, S. Lee, K. Kang and K. Park, Appl. Phys. Lett. **83**, 2611 (2003).

[2] H.C. Lin, M.F. Wang, F.J. Hou, J.T. Liu, T.Y. Huang and S.M. Sze, J. J. Appl. Phys. **41**, L626 (2002).

[3] J.A. Knapp, S.T. Picraux, C.S. Wu and S.S. Lau, J. Appl. Phys. **58**, 3747 (1985).

[4] R.D. Thompson and K.N. Tu, Thin Solid Films **93**, 265 (1982).

[5] A. Travlos, N. Salamouras, and N. Boukos, J. Phys. Chem. Solids **64**, 87 (2003).

[6] G.H. Shen, J.C. Chen, C.H. Lou, S.L. Cheng, and L.J. Chen, J. Appl. Phys. **84**, 3630 (1998).

[7] W.C. Tsai, K.S. Chi, and L.J. Chen, J. Appl. Phys. **96**, 5353 (2004).

[8] C.S. Wu, S.S. Lau, T.F. Kuech, and B.X. Liu, Thin Solid Films **104**, 175 (1983).

[9] S.S. Lau, C.S. Pai, C.S. Wu, T.F. Kuech, and B.X. Liu, Appl. Phys. Lett. **41**, 77 (1982).

[10] C.H. Luo and L.J. Chen, J. Appl. Phys. **82**, 3808 (1997).

[11] A. Travlos, N. Salamouras, and E. Flouda, Appl. Surf. Sci. **120**, 355 (1997).

[12] Y.K. Lee, N. Fujimura, T. Ho, and N. Hoh, J. Crys. Growth **134**, 247 (1993).

[13] N. Frangis, J. Van Landuyt, G. Kaltsas, A. Travlos, and A.G. Nassiopoulos, J. Crys. Growth **172**, 175 (1997).

[14] J.E. Baglin, F.M. D'heurle and C.S. Petersson, J. Appl. Phys. **52**, 2841 (1981).

[15] J.Y. Veuillen, C. d'Anterroches and T.A. Nguyen Tan, J. Appl. Phys. **75**, 223 (1994).

**Figure captions:**

1. X-ray diffraction GADDS (general area detector diffraction system) scans of an erbium silicide film formed (a) on a Si(001) substrate and (b) on a preamorphized silicon substrate and annealed at 500°C for 60 s. The x-axes are 2θ and the y-axes are χ. The insets are pole figures along the ErSi$_{2-x}$ $\{1\bar{1}00\}$ direction.

2. (a) Optical image of pyramid structural defects in an erbium silicide film annealed at 500°C for 60 s. (b) Atomic force micrograph showing the pyramidal shape of one defect. (c) A linear scan through along the pyramidal defect (through the peak) in (b) showing the shape and height profile more clearly.

3. Cross-sectional TEM micrographs of pyramidal structural defects on an erbium silicide film annealed at 500°C for 60 s. Figure (a) shows a buckled region that is completely filled with Si, while (b) shows a void at the center of the defect that extends from film to interface. The insets to (b) show a HRTEM image and the corresponding electron diffraction pattern of the defective Si region between the buckled film and the original interface.

4. Schematic representation of the formation and evolution of the pyramidal defect, showing (a) the initial, compressively stressed ErSi$_{2-x}$ film on a Si substrate, (b) the buckling of the silicide film and its separation from the substrate, (c) Si atom diffusion along Si(001) free surface and within the ErSi$_{2-x}$ film, (d) a partially filled pyramidal defect as in Fig. 3(b), (e) a completely filled defect as in Fig. 3(a), and (f) the formation of a crack in the ErSi$_{2-x}$ film in the pyramidal defect upon cooling.

5. The average size of the defects formed at 600°C as a function of annealing time. The zero of time corresponds to the beginning of the rapid thermal anneal and the first data point is at one second.

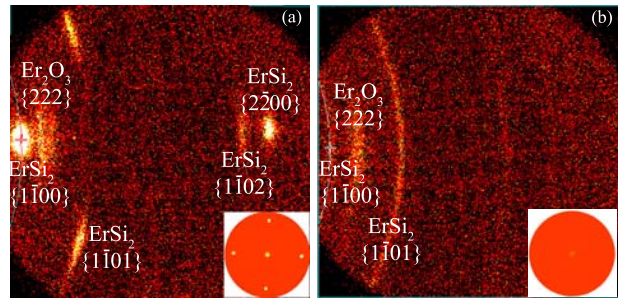

Fig. 1

E.J. Tan *et al.*

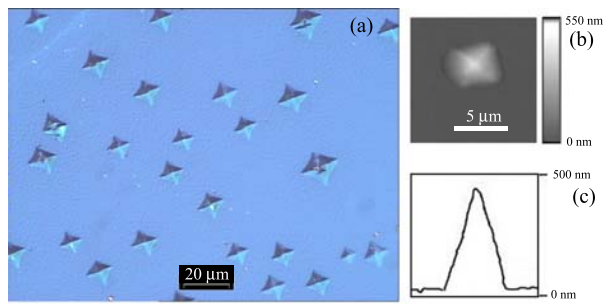

Fig. 2

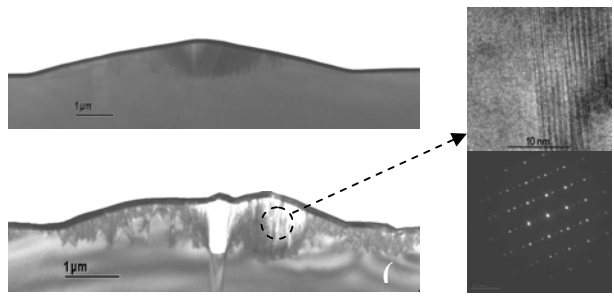

Fig. 3

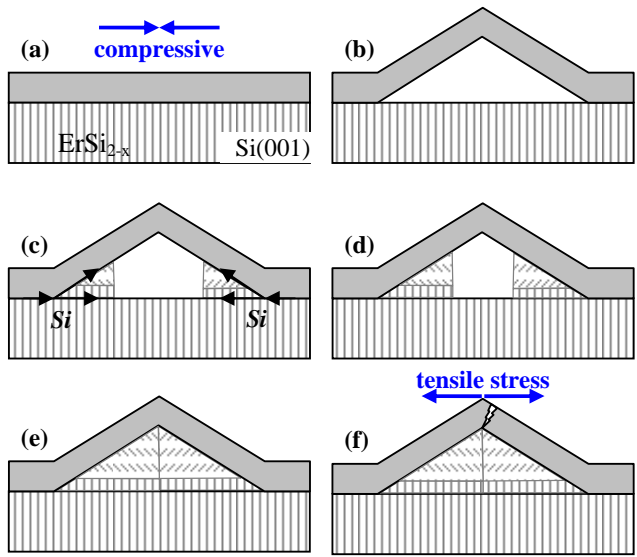

Fig. 4

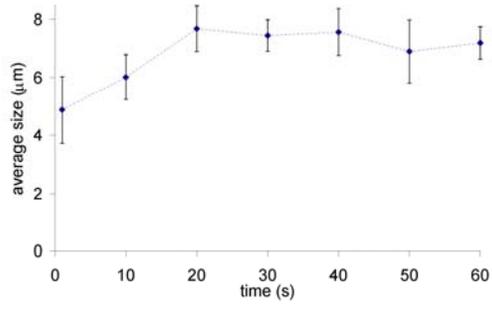

Fig. 5

E.J. Tan *et al.*